\documentclass[conference]{IEEEtran}
\IEEEoverridecommandlockouts
\usepackage{cite}
\usepackage{amsmath,amssymb,amsfonts}
\usepackage{algorithmic}
\usepackage{graphicx}
\usepackage{textcomp}
\usepackage{xcolor}
\usepackage{url,times}
\usepackage{color}
\usepackage{hyperref}
\usepackage{multirow, boldline}
\usepackage{xcolor, soul}

\usepackage{subcaption} % for subfigure

\usepackage{pdfpages}   % for inlcude pdf

\usepackage[export]{adjustbox}

\def\BibTeX{{\rm B\kern-.05em{\sc i\kern-.025em b}\kern-.08em
    T\kern-.1667em\lower.7ex\hbox{E}\kern-.125emX}}
\begin{document}

% \title{Conference Paper Title*\\

\title{Individualized Conditioning and Negative Distances for Speaker Separation}
  
\author{ \IEEEauthorblockN{Tao Sun\IEEEauthorrefmark{1}, Nidal
    Abuhajar\IEEEauthorrefmark{1}, Shuyu Gong\IEEEauthorrefmark{1},
    Zhewei Wang\IEEEauthorrefmark{4}, Charles
    D. Smith\IEEEauthorrefmark{2},
    Xianhui Wang\IEEEauthorrefmark{3},
    Li Xu\IEEEauthorrefmark{3}, Jundong Liu\IEEEauthorrefmark{1}}
  \\
  \IEEEauthorblockA{\IEEEauthorrefmark{1}School of Electrical
    Engineering and Computer Science, Ohio University, Athens, OH
    45701} \IEEEauthorblockA{\IEEEauthorrefmark{2}Department of
    Neurology, University of Kentucky, Lexington, KY 40536}
  \IEEEauthorblockA{\IEEEauthorrefmark{3}Division of Communication
    Sciences, Ohio University, Athens, OH 45701}
  \IEEEauthorblockA{\IEEEauthorrefmark{4} Massachusetts General Hospital, Boston, MA 02114}}

\maketitle

\begin{abstract}

%The ultimate goal of speech enhancement is to improve speech
%quality and intelligibility. Integrating human speech elements into
%waveform denoising neural networks has proven to be a simple yet effective
%strategy for this purpose.  Such integration, however, has mostly been
%carried out within supervised learning settings, without
%taking advantage of the power of the latest self-supervised learning
%models, which have demonstrated remarkable capability of extracting
%knowledge from large training sets.
  
%In this paper, we present {\it K-SENet}, a knowledge-assisted
%waveform framework for speech enhancement. Wave-U-Net is utilized as the
%baseline model and the foundation to build our framework. To achieve
%enhanced intelligibility, we propose a perceptual loss function that
%relies on self-supervised speech representations pretrained on large
%datasets, to provide guidance for the baseline network.  Wav2vec and
%PASE are the choices of self-supervised models in this work.

%Our proposed perceptual loss is calculated upon the perceptual
%similarities captured by the speech representations. Minimizing this
%loss would ensure the denoised network outputs sound like clean human
%speeches. Experiments on the Noisy VCTK and modified TIMIT datasets
%demonstrate that our K-SENet can significantly improve the perceptual
%quality of network outputs.

Speaker separation aims to extract multiple voices from a mixed
signal.  In this paper, we propose two speaker-aware designs to
improve the existing speaker separation solutions. The first model is
a {\it speaker conditioning} network that integrates speech samples to
generate individualized speaker conditions, which then provide
informed guidance for a separation module to produce well-separated
outputs.

The second design aims to reduce non-target voices in the separated
speech. To this end, we propose {\it negative distances} to penalize
the appearance of any non-target voice in the channel outputs, and
{\it positive distances} to drive the separated voices closer to the
clean targets.
% An combined {\it auxiliary loss} is introduced to integrate negative
% disatnce with {\it positive distances} that drive the separated
% voices close to clean target voices.
We explore two different setups, {\it weighted-sum} and {\it
  triplet-like}, to integrate these two distances to form a combined
auxiliary loss for the separation networks. 
% loss.
Experiments conducted on LibriMix demonstrate the effectiveness of our
proposed models.
% We also conduct an ablation study to inspect the contirubtions of
% the negative term.
  
% The ultimate goal of speech enhancement (SE) is to improve speech
% quality and intelligibility.  Integrating human speech components into
% denoising neural networks has proven to be a simple yet effective
% strategy for this purpose. So far, such integration, however, has been
% mostly carried out within the same training datasets, without taking
% advantage of some large-scale pretrained speech networks which can
% provide insights and knowledge to characterize human speeches.

% In this paper, we present K-SENet, a knowledge-assisted
% framework for speech enhancement. Our baseline SE network is built as
% a fully convolutional network (FCN). To achieve enhanced
% intelligibility, we propose a perceptual loss function that relies on
% speech representations pretrained on large datasets, to provide
% guidance for the baseline network. This new loss is calculated upon
% the perceptual similarities captured by speech representations.
% % , instead of point-wise differences in the temporal space.
% Minimizing this loss would ensure our FCN outputs {\it sound like}
% clean human speeches, as in the training data of the pretrained models. 

% To the best of our knowledge, this is the first work that attempts to
% integrate large-scale pretrained knowledge networks into SE solutions.
% Experiments on the VCTK and TIMIT datasets demonstrate that our
% K-SENet can significantly improve the perceptual quality of
% network outputs.

\end{abstract}


@inproceedings{
  
}

% Content from Tao's disseration
%
% speech separation (p14, para1)
% speaker separation (p14, para2)
% area overview: speech separation (page 15, para1)
% speaker separation (1.1.2)
% proposal & contributions: page 19, para 2, 3
% these two are independent, right? 

\section{Introduction}
\label{ch:intro}

% speech separation (p14, para1)

Speech separation, also known as the cocktail party problem,
% is a special case of source separation. It
aims to separate a target speech from its background interference
\cite{wang2018supervised}. It often serves as a preprocessing step in
real-world speech processing systems, including ASR, speaker
recognition, hearing prostheses, and mobile telecommunications.
% Speech separation
{\it Speaker separation} (SS) is a sub-problem of speech separation,
where the main goal is to extract multiple voices from a mixed signal.

%While the human auditory
%system has extraordinary speech separation capabilities, designing
%artificial models for this function has proven to be very challenging
%\citep{wang2018supervised}.
%
% speaker separation (p14, para2)
%
% In monophonic speech separation (i.e., single microphone), background
% interference includes interfering speech, background noise, and room
% reverberation, corresponding to its three sub-regions (speaker
% separation, speech enhancement (SE), and speech dereverberation),
% respectively.
%This paper focuses on the speaker separation problme:
%\textit{speaker separation} separates multiple voices from a mixed
%signal.
%
% In this paper, we are particularly interested in the {\it speaker
%   separation} problem: separating multiple voices froma mixed signal.

%
% area overview: speech separation (page 15, para1)
%With the arrival of the deep learning wave, speech separation
%solutions have entered the era of deep neural networks (DNNs). Despite
%the emergence of a large number of excellent DNN models for speech
%enhancement and speaker separation, the existing solutions for speech
%separation are not perfect and there is still room for improvement.

%In this section, we briefly introduce traditional and
%DNN-based solutions for speech enhancement and speaker separation and
%their evolution. Especially, some issues in current time-domain DNN
%solutions for these two tasks are identified.

% speaker separation (1.1.2)

%\subsection{Speaker Separation Solutions}

% Mimicing the mechanism of human auditory systems,
Following the mechanism of the human auditory system, traditional
speaker separation solutions commonly rely on certain heuristic
grouping rules, such as periodicity and pitch trajectories, to
separate mixed signals
\cite{wang2006computational,hu2006auditory,hu2010tandem}.
% They typically compute a two-dimensional ideal time-frequency (T-F)
% mask \cite{wang2006computational} for each speech source based on
% these rules, and then apply them to the time-frequency representation
% of the mixed signal. Due to the hand-engineered essence of the
% grouping rules, performance of these methods is limited
% \cite{wang2013towards}.
%
Two-dimensional ideal time-frequency (T-F) masks are often generated
based on these rules and applied to
% used to model and extract different speech sources, which are then
% applied to
mixed signals to extract individual sources. Due to the hand-crafted
nature, these grouping rules, however, often have limited
generalization capability in handling diverse real-world signals
\cite{wang2013towards}.

Similar to many other AI-related areas, deep neural networks (DNNs)
have recently emerged as a dominant paradigm to solve the SS
problems. Early DNNs were mostly frequency-domain models
\cite{isik2016single,yu2017permutation,kolbaek2017multitalker,chen2017deep,wang2018alternative},
aiming to approximate ideal T-F masks and rely on them to restore
individual sources through short-time Fourier transform (STFT). As the
modified T-F representations may not be converted back to the time
domain,
% This has motivated researcher to develop waveform DNN models in recent years.
these methods commonly suffer from the so-called {\it invalid STFT
  problem} \cite{luo2018tasnet}.

Waveform-based DNN models have grown in popularity in recent
years,
%in part due to their capability to
partly because they can avoid the invalid STFT problem
\cite{luo2018tasnet,gong2019dilated, luo2019conv, abuhajar2021network,
  sun2021boosting}.
% the past three years or so.
Pioneered by TasNet \cite{luo2018tasnet} and Conv-TasNet
\cite{luo2019conv}, early waveform solutions tackle the separation
task with three stages: encoding, separating, and decoding. However,
speaker information is often not explicitly integrated into the
network training and/or inference procedures.

Speaker-aware SS models
\cite{wang2021dual,shi2020speaker,zeghidour2021wavesplit,shi2020speech,liu2021permutation,nachmani2020voice}
%aiming to incorporate speaker information into the separation process.
provide a remedy in this regard. This group of solutions can be
roughly divided into {\it speaker-conditioned} methods
\cite{wang2021dual,shi2020speaker,zeghidour2021wavesplit} and {\it
  auxiliary-loss} based methods
\cite{shi2020speech,liu2021permutation,nachmani2020voice}.  The former
% \cite{wang2021dual,shi2020speaker,zeghidour2021wavesplit}
% commonly consist of a speaker module and a separation module. The
rely on a speaker module to infer speaker information, which is then
taken as conditions by a separation module to generate separated
output waveforms. The existing speaker-conditioned solutions, however,
are either not in the time-domain
\cite{wang2021dual,shi2020speaker,zeghidour2021wavesplit} or do not
explicitly integrate speech information into the speaker conditioning
process \cite{zeghidour2021wavesplit}.

Auxiliary-loss based methods
\cite{shi2020speech,liu2021permutation,nachmani2020voice}
% train DNN models using composite loss functions to separate waveform
% mixtures.
achieve speaker awareness through composite loss functions.  In
addition to a main loss, an auxiliary loss (or losses) is used to
incorporate speakers' information into the network training procedure.
%In addition, certain auxiliary losses are used to incorporate speaker
%information into their models, leading to improved network
%performance.
Such auxiliary losses are commonly formulated to ensure a match
between network outputs and the target speakers. However, to the best
of our knowledge, no solution has attempted to explicitly suppress
%information
voices from other non-target speakers. As a result, residual voices of
non-target speakers are often noticeable in the network outputs.

% proposal & contributions: page 19, para 2, 3

In this paper, we propose two waveform speaker separation models to
address the aforementioned limitations. The first model is a {\it
  speaker conditioning} network that integrates individual speech
samples in the speaker module to produce tailored speaker conditions.
The integration is based on speaker embeddings computed through a
pretrained speaker recognition network.
% The second solution is designed to suppress non-target speaker
% voices thoroughly in the separated speech.
The second solution aims to completely suppress non-target speaker
voices in the separated speech.  We propose an {\it auxiliary loss}
with two terms: the first drives the separated voices close to target
clean voices, while the second term penalizes the appearance of any
non-target voice in the separated outputs.  The latter, which we call
{\it negative distances}, is achieved by maximizing distances between
the speech representations of extracted sources and those of the
non-target sources. We also explore different schemes to integrate the
proposed distances.
%
%Cleaner signal is produced 
%
%Specifically, the auxiliary loss in the
%proposed solution has two terms: the first drives separated voices to
%get close to target clean voices, and we use the second term to
%suppress the voices of the non-target speakers. Such voice suppression
%is achieved by maximizing distances between speech representations of
%extracted sources and each of clean non-target sources.

\section{Background}

%In this section, background knowledge concerning our proposed
%speaker-aware
% speaker separation
%SS models is briefly explained, which includes conditioning in machine
%learning, triplet loss, and speech representations generated through
%self-supervised learning.

We start with this section to provide some background knowledge
concerning our proposed speaker-aware SS models, which includes
conditioning in machine learning, triplet loss, and speech
representations generated through self-supervised learning.

%
%the permutation problem. While conditioning and
% the triplet loss are respectively relevant to the proposed
% speaker-conditioned solution and auxiliary-loss based solutions, the
% permutation problem is a common problem that occurs in almost all
% DNN-based speaker separation solutions.

%\subsection{FiLM: A Conditioning Implementation}
\subsection{Conditioning and FiLM}
% https://distill.pub/2018/feature-wise-transformations/

In everyday life, it is often helpful to process one source of
information in the context of another. For example, video and audio in
a movie can be better understood in the context of each other. This
context-based processing is called {\it conditioning} in machine
learning, where computations through a model are conditioned or
modulated by information extracted from auxiliary inputs.
For speaker-conditioned speaker separation, conditioning in a network
can be done through its separation module, which would take speaker
information as the context to produces the output voices.
% on the speaker information inferred by its
% speaker module.

% Neural network conditioning is the criteria to provide a condition to
% the network to aid learning based on the provided condition. In our
% proposed work, we manage the task of speaker separation by
% conditioning the network on speakers' information that is extracted
% from speech. This criteria has shown a significant improvement over
% non-conditioned networks.

\begin{figure}
  \includegraphics[width=9cm]{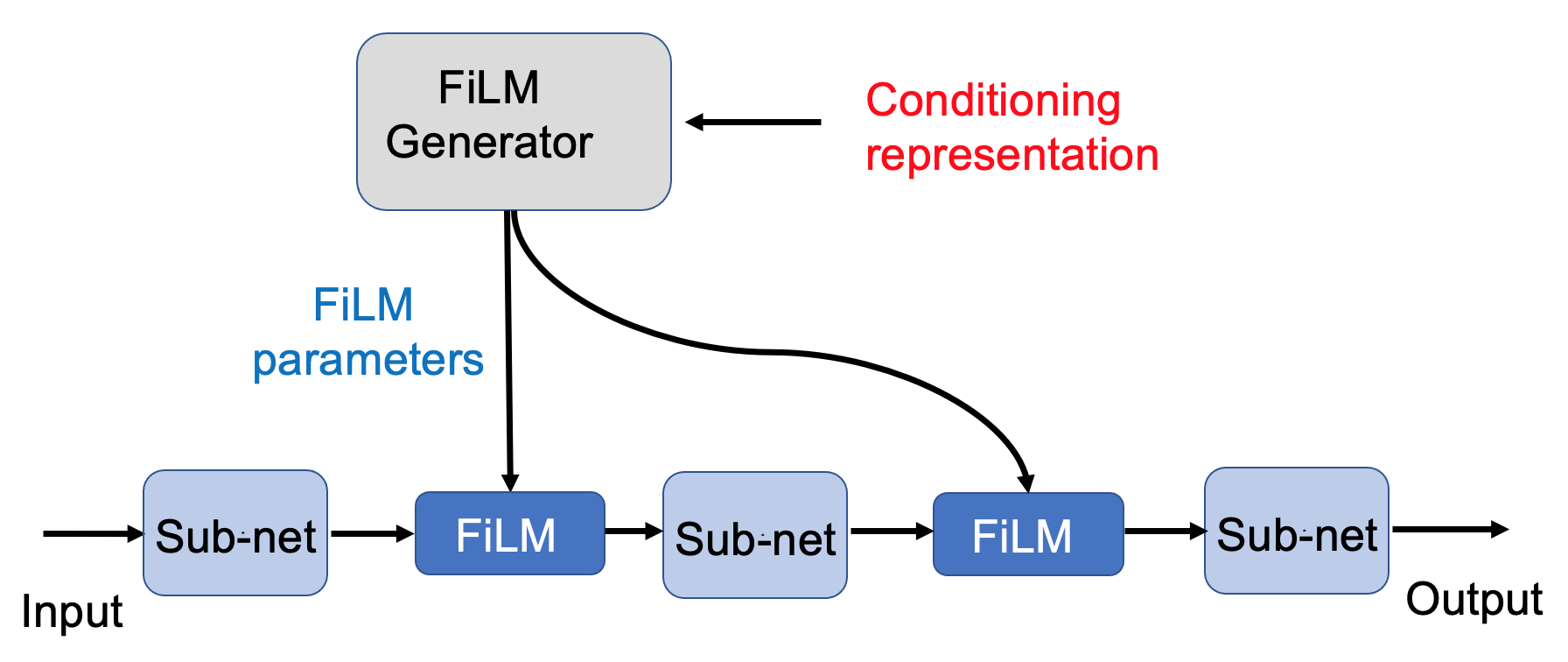}
    \caption{Feature-wise Linear Modulation (FiLM) architecture. }
    \label{fig:FiLM2}
\end{figure}

{\it Feature-wise Linear modulation} (FiLM) \cite{perez2018film} is a
popular feature conditioning method that was shown to enhance the
performance of
% neural networks
% based on {\bf arbitrary??}
% conditions. FilM has been integrated in various
neural network solutions for a variety of tasks, including visual
reasoning and speech separation \cite{zeghidour2021wavesplit,
  wang2021dual}. FiLM learns to adaptively influence the output of a
neural network by applying an affine transformation to the network’s
intermediate features, based on some input. As shown in
Fig.\ref{fig:FiLM2}, FiLM conditioning architecture consists of a FiLM
generator and one or more FiLM layers.
The generator takes a conditioning representation as input and
generates FiLM vector parameters, which are later used in the FiLM
layers to modulate the input with an affine transformation, i.e., a
combination of conditional biasing and conditional scaling.

%Let $\mathbf{x}$ and $\mathbf{z}$ denote inputs of a FiLM layer and
%conditioning representations respectively, the  layer output
%can be written as:
%\begin{equation}
% \textrm{FiLM}(\mathbf{x})= \gamma(\mathbf{z})\odot
% \mathbf{x}+\beta(\mathbf{z})
% % \textrm{FiLM}(x)= \gamma(z)\odot x+\beta(z)}
%\end{equation}
%where $\odot$ is element-wise product; $\gamma$ and $\beta$ are
%scaling and shifting functions (vectors) learned by FiLM.

\subsection{Triplet Loss}
In machine learning, we often consider triplet samples \cite{schroff2015facenet},
each consisting of an anchor input, a matching input with the same
label (called a positive sample), and a non-matching input with a
different label (called a negative sample).  The triplet loss,
initially introduced in metric learning \cite{schultz2004learning}, is
a loss function based on relative comparisons, i.e., an anchor $x^a$
is compared to one positive sample $x^p$ and one negative sample
$x^n$.

\begin{figure}[h]
  \includegraphics[width=9cm]{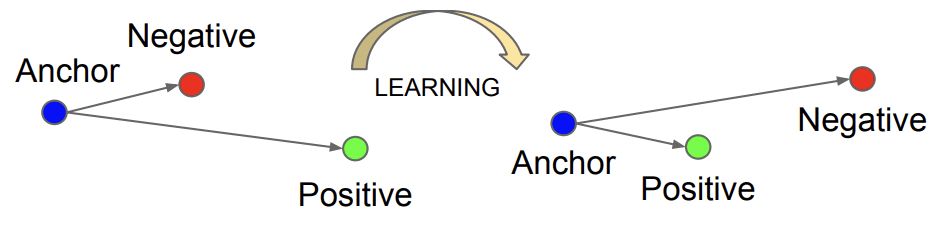}
    \caption{Learning through the triplet loss \cite{schroff2015facenet}. }
    \label{fig:triplet}
\end{figure}

Shown in Fig. \ref{fig:triplet}, a triplet loss learns embeddings to
minimize the distance between an anchor input and the positive
samples, and at the same time maximize the distance from the anchor to
the negative inputs. More specifically, for one triplet, we want:
\begin{equation}
  ||f(x^a) - f(x^p)||_2^2 + \alpha < ||f(x^a) - f(x^n)||_2^2\,,
\end{equation}
where $f(\cdot)$ is the embedding function and $\alpha$ is defined as
the minimum margin between positive and negative pairs.
% If the set of
% all possible triplets $(f(x_i^a), f(x_i^p), f(x_i^n))$ has cardinality
% $N$, the triplet loss can be written as
% \begin{equation}
%   L = \frac{1}{N}\sum_i^N \textrm{max}(0, ||f(x_i^a) - f(x_i^p)||_2^2
%  - ||f(x_i^a) - f(x_i^n)||_2^2 + \alpha)\,.
% \end{equation} 

%\subsection{Self-Supervised Learning}
%Self-Supervised Learning {\bf(SSL)} is a deep learning paradigm that
%is widely used in numerous machine learning areas such as Natural
%Language Processing {\bf(NLP)[cite word2vec]}, Computer Vision
%{\bf(CV)}, and speech related tasks. It is primarily trained on
%unlabeled data to learn meaningful feature representations from the
%data, which is considered an advantage due to the lack of labeled data
%for certain tasks.

%{\bf SSL} is commonly used for downstream tasks. For example, Word2Vec
%{\bf cite} is trained on huge corpus to learn words/sentences
%representations that are later used in various downstream tasks in
%NLP, and it has shown a significant improvement compared to using raw
%representations. Furthermore, Wav2Vec{\bf[cite]} and PASE {\bf[cite]}
%are other examples of SSL models that are trained on long hours of
%speech data to be used in many speech related tasks such as speech
%enhancement, speech verification, speech recognition, etc.

\subsection{Speech Representations via Self-supervised Learning} 

In self-supervised learning (SSL), models are trained to predict one
part of the data from other parts \cite{liu2020self}. SSL models for
speech data and tasks commonly aim to output speech representations in
the form of compact vectors that capture high-level semantic
information from raw speech data \cite{pascual2019learning,
  Schneider_2019_wav2vec, ravanelli2020multi, liu2020tera,
  baevski2019vq}. In our work, we utilize the speech feature
representations generated from Wav2vec\cite{Schneider_2019_wav2vec},
an SSL network, to pass high-level meaningful features to our
networks.

Wav2vec is trained on LibriSpeech corpus using the contrastive
predictive coding (CPC) loss\cite{baevski2019vq} to pretrain speech
representations for ASR tasks. Experiment results showed that wav2vec
can significantly improve the performance over the chosen baseline
solutions.
%
% Wav2vec consists of two parts; an encoder and a context
% network. The former has seven-layer convolutional network, and its
% duty is to extract latent features from the inputs. Later, the context
% network combines multiple outputs from the encoder into a
% contextualized tensor, where to be used later for downstream tasks.
%
Wav2vec consists of two parts, an encoder and a context network. The
former is a seven-layer convolutional network, and its functionality
is to extract latent features from the inputs. The context network
combines multiple outputs from the encoder into a contextualized
tensor, which then could be used for downstream tasks.

\section{Method}

%Let $\mathbf{x}_n$ be a waveform produced by mixing $C$ sources
%$\mathbf{x}_1, \mathbf{x}_2, ..., \mathbf{x}_c$, i.e.,
%
%\begin{equation}
%  \mathbf{x}_n = \sum_{i=1}^{C} \mathbf{x}_i
%\end{equation}

Let $\mathbf{X}$ be a waveform produced by mixing $C$ sources
$\mathbf{x}_1, \mathbf{x}_2, ..., \mathbf{x}_c$, i.e.,

\begin{equation}
  \mathbf{X} = \sum_{i=1}^{C} \mathbf{x}_i
\end{equation}

%A waveform monaural speaker separation model aims to directly separate
%the mixed signal $\mathbf{x}_n$ into $C$ estimations
%$\tilde{\mathbf{x}_1}, \tilde{\mathbf{x}_2}, ...,
%\tilde{\mathbf{x}_c}$.

A waveform monaural speaker separation model aims to directly separate
the mixed signal $\mathbf{X}$ into $C$ estimations
$\tilde{\mathbf{x}_1}, \tilde{\mathbf{x}_2}, ...,
\tilde{\mathbf{x}_c}$.

\subsection{Baseline Model}

Conv-TasNet \cite{luo2019conv}, a variation of
TasNet\cite{luo2018tasnet} is adopted as the baseline model
% in our experiments.
in this work.  Conv-TasNet has been shown to generate
% superior
remarkable results for many speech tasks including speech separation.
% As many speech separation solutions,
Inspired by the T-F domain masking-based speech separation solutions,
Conv-TasNet follows an encoder-separator-decoder architecture.
% Better than its former TasNet, Conv-TasNet resolves the issue of
% using Bi-directional Long Short Term Memory (BLSTM) in their decoder
% that leads to losing long dependencies across long sequences.
The encoder converts a waveform mixture into a feature map by a linear
transformation that emulates STFT. The separator is a network which
learns the masks for each source in the mixed inputs. After the
learned masks are applied to the encoder output, the results are fed
into the decoder, which also carries out a linear transformation. The
decoder emulates the inverse STFT (iSTFT) operation to compute the
final separated outputs.

% The encoder of Conv-TasNet maps the input signal to high level feature
% representations that flows into the separator where in its turn learns
% separation masks for each target source. Later the separated sources
% are generated by multiplying the masks with the input mixture.

% \vspace{\baselineskip}
%\begin{figure}[htb]
\begin{figure*}
   \centering
   \includegraphics[width=13cm]{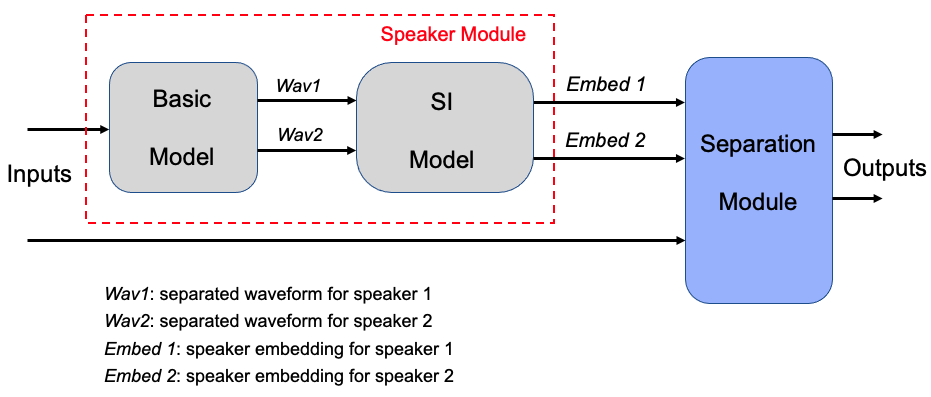}
   \caption{Architecture of the proposed speaker-conditioned
     solution. Refer to text for details. }
    \label{fig:twostage}
\end{figure*}

\subsection{Proposed Speaker-conditioned Model}
The architecture of our proposed speaker-conditioned pipeline is shown
in Fig~\ref{fig:twostage}. It consists of two major modules, a {\it
  speaker module} (red box), followed by a {\it separation
  module}. The speaker module infers the speaker information in the
input audio samples, which is then sent as {\it conditions} to the
separation module to generate outputs.

In the speaker module, a {\it basic model} is first used to infer
intermediate separated sources from input mixture waveforms.  These
separated sources, illustrated as {\it Wav1} and {\it Wav2} in
Fig.~\ref{fig:twostage}, are then fed into a pretrained speaker
identification (SI) network to generate respective speaker embeddings
of the separated sources ({\it Embed 1} and {\it Embed 2} in
Fig.~\ref{fig:twostage}).
%and they are used as
%conditions in the separation module to produce the final separated
%outputs.
Speaker embeddings generated this way contain not only the collective
information about the speaker, but also individualized details of each
speech sample.
%input into a pretrained speaker identification (SI) model to extract
%speaker embeddings as conditions. Such speaker embeddings are not only
%specific to an speaker, but also highly correlated to the input
%speeches.
As a result, rich and better customized speech information is
integrated.

In this work, Conv-TasNet and RawNet2 \cite{jung2020improved} (a
pretrained waveform speaker verification model) are used as the basic
model and SI model, respectively.  Note the choices are not unique --
for example, speaker separation models such as Dual-path methods
\cite{luo2020dual,chen2021dual,subakan2021attention} could also be
used as the basic model.

In the separation module of our proposed pipeline, the input mixture
signal is modulated by FiLM parameters generated through the speaker
embeddings, prior to being mapped into final separated results. The
implementation of our separation module is the same as that in
Wavesplit \cite{zeghidour2021wavesplit}, which has 40 dilated causal
convolution layers.  Among these 40 layers, every 10 layers are put
into a group. The dilation rate of the first layer of each group is
set to 1, and the subsequent dilations increase sequentially.
% As mentioned in Chapter \ref{ch:bg}, such layers are also used in
% the TCN and Conv-TasNet.
A similar layer setup has also been used in Conv-TasNet.

It is worthy to note that our proposed speaker-conditioned pipeline
can be further refined by extending it into a recurrent model:
% , as illustrated in Fig.~\ref{fig:recurtwostage}(b).
final separated outputs of the separation module can be taken as
intermediate separated sources to the SI model to generate refined
speaker embeddings, which are then inserted as conditions into the
separation module for the next iteration.

\subsection{Proposed Auxiliary-loss based Models}
\begin{figure*}[htb]
   \centering
   \includegraphics[width=15cm]{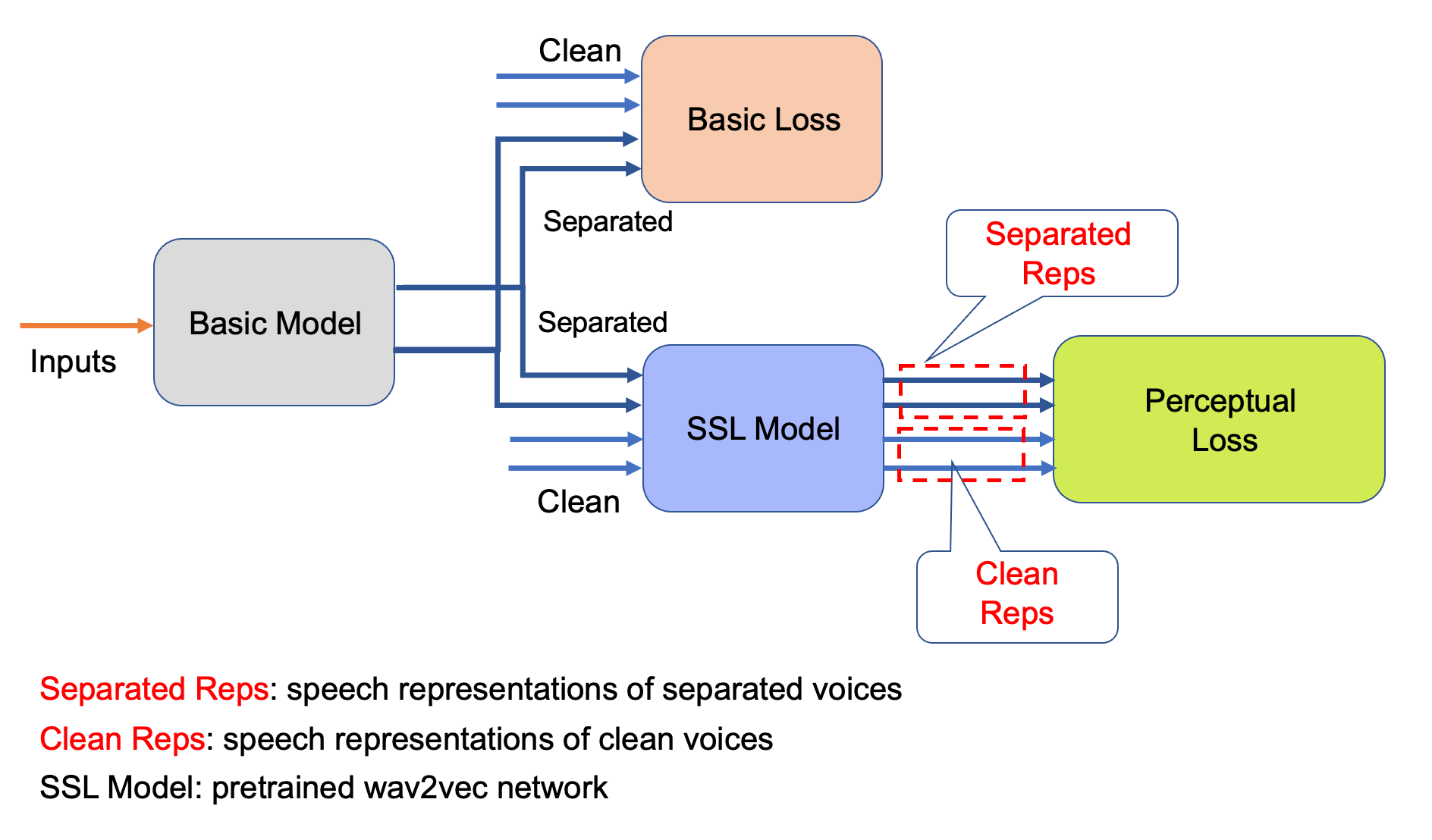}
    \caption{The architecture of our proposed auxiliary-loss based solution.}
    \label{fig:aux_loss}
\end{figure*}

Fig. \ref{fig:aux_loss} shows the architecture of our proposed
auxiliary-loss based solutions, where we take two-speaker
% mixtures as the testbed.
separation as an example to illustrate our design.  Mixtures are fed
into a basic model, which is Conv-TasNet \cite{luo2019conv} in this
work. The basic model is trained with the SI-SNR loss to produce
separated sources $\tilde{\mathbf{x}_1}$ and $\tilde{\mathbf{x}_2}$.
% As mentioned above, speech representations
% out of SSL models are taken as speaker embeddings in our proposed
% solutions.
In this work, we use speech representations
% based on
generated through wav2vec\cite{Schneider_2019_wav2vec}, a pretrained
self-supervised learning model, as speaker embeddings in our proposed
framework.

Most existing auxiliary-loss based solutions are designed to ensure
separated voices sound like target speakers, which can be achieved by
minimizing the dissimilarities (or distances) between speaker
embeddings of predicted sources and clean (ground-truth) target
sources.
% Such distances can be called \textit{positive distances}, which is
% formally written as:
We call these distances \textit{attraction} or \textit{positive
  distances}, which can be written as:

%Current auxiliary-loss based solutions are designed to ensure
%separated voices sound like target speakers, which is achieved by
%minimizing the distances between speaker embeddings of predicted
%sources and clean target sources. Such distances can be called
%\textit{positive distances}, which is formally written as:

%\begin{equation}
%          l_1^{(i)} = \sum_{m}\frac{1}{|\phi_{m}(\mathbf{x}_i)|}
%          ||\phi_{m}(\mathbf{x}_i) -
%          \phi_{m}(\tilde{\mathbf{x}_i})||_{2}^{2}
%\end{equation}

\begin{equation}
         \textrm{dist}_\textrm{pos}^{(i)} = \sum_{m}\frac{1}{|\phi_{m}(\mathbf{x}_i)|}
          ||\phi_{m}(\mathbf{x}_i) -
          \phi_{m}(\tilde{\mathbf{x}_i})||_{2}^{2}
\end{equation}
where $\phi_{m}(\mathbf{x}_i)$ represents the $m^{th}$ vector in the
representation of $\mathbf{x}_i$. These solutions, however, have no
mechanism to suppress sounds of non-target speakers. As a result,
residual sounds of non-target speakers can be easily perceived in
the separated voices. %  of these solutions.
%To address this issue, we propose
%to add a new {\bf negative} term to the existing auxiliary loss
%functions that suppresses information from non-target speakers.
To address this issue, we propose to a new {\it repulsion term} to
reduce the output from non-target speakers.  More specifically, this
term reduces non-target voices by maximizing \textit{repulsion} or
\textit{negative distances}, which are defined as the distances
between speaker embeddings of the predicted source and the
ground-truth non-target sources:

\begin{equation}
          \textrm{dist}_\textrm{neg}^{(i)} = \sum_{j \neq
            i}\sum_{m}\frac{1}{|\phi_{m}(\mathbf{x}_i)|}
          ||\phi_{m}(\mathbf{x}_i) -
          \phi_{m}(\tilde{\mathbf{x}_j})||_{2}^{2}
\end{equation}      

%{\bf Two integration schemes}
%
%We introduce two methods to integrate
%both positive distances $l_1$ and negative distances $l_2$ to an
%auxiliary loss function, which we call \textit{perceptual loss}. In
%the first method, we define $L_1$ and $L2$ as
%

%
% \vspace{\baselineskip}
%\begin{figure}[htb]
%
%
% figure*, the * is crucial for one-column?
%
%

{\bf Two different integration schemes}
% Our proposed {\it
%  negative distances} $\textrm{dist}_\textrm{neg}$ can be integrated
% with the {\it positive distances} $\textrm{dist}_\textrm{pos}$ in many
%different ways.
There could be many different ways to integrate our proposed {\it
  negative distances} $\textrm{dist}_\textrm{neg}$ with the {\it
  positive distances} $\textrm{dist}_\textrm{pos}$. In this work, we
explore two setups for this task.
%
% We introduce two methods to integrate the attraction distances
% $dist_\textrm{attract}$ and repulsion distances
% $dist_\textrm{repel}$ to an auxiliary loss function, which we call
% \textit{perceptual loss}.
Both of them form an auxiliary loss function for the network, which
we call \textit{perceptual loss}.

In the first setup, we define collective distances over the entire
training set,  
%$L_1$ and $L2$ as
$D_\textrm{pos}$ and $D_\textrm{neg}$, as:

% {\bf Integration via weighted sum} In the first , we define
% $D_\textrm{attract}$ and $D_\textrm{repel}$ as

%\begin{subequations}
%\begin{align}
%          L_1 &= \frac{1}{N}\sum_{i}l_1^{(i)} \\
%          L_2 &= \frac{1}{N}\sum_{i}l_2^{(i)},
%\end{align}
%\end{subequations}

\begin{equation}
%\begin{align}
          D_\textrm{pos} =
          \frac{1}{N}\sum_{i}\textrm{dist}_\textrm{pos}^{(i)}, 
          \ \ \ \ \ D_\textrm{neg} = \frac{1}{N}\sum_{i}\textrm{dist}_\textrm{neg}^{(i)},
%\end{align}
\end{equation}
where $N$ is the number of training examples.
%\hl{Similar to the perceptual loss for SE in Equation ??},
We then define the perceptual loss as a weighted summation of
$D_\textrm{pos}$ and $D_\textrm{neg}$:

%\begin{equation}
%          L_{perc}^1 = \lambda_1 L_1+\frac{\lambda_2}{L_2},
%\end{equation}
\begin{equation}
          L_\textrm{perc}^1 = \lambda_1
          D_\textrm{pos}+\frac{\lambda_2}{D_\textrm{neg}},
          \label{L_weightedsum}
\end{equation}
where $\lambda_1$ and $\lambda_2$ are weighting coefficients, which
can be set manually or empirically in experiments.

In the second setup, we emulate a triplet loss to enforce a minimal
margin $\alpha$ between the positive distance
$\textrm{dist}_\textrm{positive}^{(i)}$ and negative distance
$\textrm{dist}_\textrm{negative}^{(i)}$ for each data sample:

%\begin{equation}
%  l_1^{(i)} + \alpha < l_2^{(i)}.
%\end{equation}

\begin{equation}
  \textrm{dist}_\textrm{pos}^{(i)} + \alpha < \textrm{dist}_\textrm{neg}^{(i)}.
\end{equation}

Thus, the perceptual loss can be defined as:
%\begin{equation}
%  L_{perc}^2 = \frac{1}{N}\sum_{i}\textrm{max}(0, l_1^{(i)} - l_2^{(i)} + \alpha),
%\end{equation}

\begin{equation}
  L_\textrm{perc}^2 = \frac{1}{N}\sum_{i}\textrm{max}(0,
  \textrm{dist}_\textrm{pos}^{(i)} - \textrm{dist}_\textrm{neg}^{(i)} +
  \alpha),
  \label{L_triplet}
\end{equation}
where $N$ is the number of training examples.

The overall loss in both setups is designed as a weighted summation of
the basic loss $L_\textrm{basic}$ (from Conv-TasNet) and the respective
perceptual loss:

%$P_\textrm{weighted-sum}$ and
%$P_\textrm{triplet-like}$ can be written as a weighted sum of the
%basic loss (from Conv-TasNet) and one of the perceptual losses:

%\begin{equation} \label{eq:comb_loss_ss}
%L = \lambda_{b} L_{\textrm{basic}}+\lambda_{p} L_{\textrm{perc}}  
%\end{equation}
\begin{equation} \label{eq:comb_loss_ss}
L = \lambda_{b} L_{\textrm{basic}}+\lambda_{p} L_{\textrm{perc}}  
\end{equation}
where $L_\textrm{perc}$ is either $L_\textrm{perc}^1$ from
Eqn.~\ref{L_weightedsum} or $L_\textrm{perc}^2$ from
Eqn.~\ref{L_triplet}. $\lambda_b$ and $\lambda_p$ are weighting
coefficients to decide the contributions, which can be set manually or
empirically in experiments.  We name the model using
$L_\textrm{perc}^1$ as $P_\textrm{weighted-sum}$ and that using
$L_\textrm{perc}^2$ as $P_\textrm{triplet-like}$.

\section{Experiments and Results}
% In this section,
% the designed experiments for our proposed models are
% described and their results are analyzed.
In this section, we conduct experiments to evaluate the effectiveness
of our proposed models.  First, we introduce the
%LibriMix
dataset used in the experiments, followed by the training strategy and
evaluation metrics for the competing models. Then, we report the
results of the three proposed speaker-aware solutions and compare them
with the baseline model. Finally, we conduct an ablation study on
$P_\textrm{weighted-sum}$, one of the proposed perceptual-loss based
solutions, to analyze the effects of the contributing terms.

\subsection{Data and Training}
LibriMix \cite{cosentino2020librimix} is an open-source dataset for
single-channel speech separation. The utterances in the mixtures of
LibriMix are taken from LibriSpeech \cite{Librispeech_2015_C}.
% LibriMix has two main datasets, Libri2Mix
% and Libri3Mix, which contain two-speaker and three-speaker mixtures,
% respectively. Each of these two datasets has two conditions, clean or
% noisy. The noise samples are taken from WHAM! \cite{wichern19wham}.
% Corresponding to the structure of LibriSpeech,
% each condition has two training sets (train-360 and train-100), one
% validation set, and one test set. Each of the sets has four variants,
% which are combinations of two sampling rates (8 kHz and 16 kHz) and
% two modes (\textit{max} and \textit{min}). While a mixture in the
% \textit{min} mode is clipped to the length of the shortest utterance,
% in the \textit{max} mode, the shorter utterance is zero-padded to the
% longer one.
%
All our models are trained for 200 epochs. During training, the
mixtures in the train-100 of clean Libri2Mix (\textit{min} mode, 16
kHz) are divided into 3-second segments as the training set. The
optimizer is Adam \cite{kingma2014adam} with the learning rate 0.001
and early stopping patience 30. We apply the utterance-level
permutation invariant training (uPIT) \cite{kolbaek2017multitalker}
with the label assignments evaluated by SI-SNR values to train the
Conv-TasNet in our proposed models.

Objective performance for speech separation can be evaluated by
metrics concerning signal fidelity (e.g., SNR and SI-SNR) and output
perceptual quality (e.g., PESQ \cite{rix2001perceptual} and STOI
\cite{taal2011algorithm}).
% Another class of objective metrics,
% including word error rate (WER) and word accuracy rate (WAR), use
% speech recognition accuracy to indirectly evaluate speech separation
% models \cite{luo2020dual,zhang2018dl}. Such approach usually takes
% speech separation models as front-end modules in tandem audio
% processing systems \cite{luo2019conv}.
In this work, we choose SI-SNR and STOI as the objective metrics to
evaluate our models.
%{\bf Please refer to Chapter \ref{ch:se} for the
%  detailed definition of SI-SNR} and its advantage over SNR as the
%object evaluation metric for speech separation.

\subsection{Results and Analysis}

% each model and its configuration and parameters. Then comparison 

As we mentioned in the Method section, Conv-TasNet is taken as the
basic model for our proposed models.
In our perceptual-loss based model $P_\textrm{weighted-sum}$, the
weights of the combined loss ($\lambda_b$ and $\lambda_p$) are
empirically set to 1.0 and 1.0, and the weights of the positive and
negative terms ($\lambda_1$ and $\lambda_2$ in
Eqn.~\ref{L_weightedsum}) are set to 100 and 0.001.  In the
triplet-loss based model $P_\textrm{triplet-like}$, the weights of the
combined loss ($\lambda_b$ and $\lambda_p$) are set to 1 and 300, and
the margin $\alpha$ is set to 0.0035.
% These hyper-parameters are set empirically.

%As we mentioned in the Method section, Conv-TasNet is taken as the
%basic model for our proposed models.  In our perceptual-loss based
%model, the weights of the combined loss ($\lambda_b$ and $\lambda_p$)
%are empirically set to 1.0 and 1.0, and the weights of the perceptual
%loss ($\lambda_1$ and $\lambda_2$) are set to 100 and 0.001. In the
%triplet-loss based model, the weights of the combined loss
%($\lambda_b$ and $\lambda_p$) are set to 1 and 300, and the margin
%$\alpha$ is set to 0.0035.
%% These hyper-parameters are set empirically.

% Table \ref{T:librimix} shows the results of our proposed models on the
% LibriMix dataset.

Table \ref{T:librimix} shows the results from the competing models on
the LibriMix dataset. The first line is the results from Conv-TasNet.
The second and third lines show the results from our perceptual-loss
models using weighted-sum and triplet-like, respectively.  The fourth
line is for our proposed model using speaker-conditioning.  It is
evident that all the three proposed models outperform the baseline
models in terms of the evaluation metrics, especially in the SI-SNR
where the performance gains are more prominent. Among the three
proposed models, the conditioning-based solution achieves the best
performance.  The performance of the two auxiliary-loss based
solutions is comparable, while $P_\textrm{weighted-sum}$, the model
with the weighted-sum perceptual loss, performs slightly better.

The demonstrated advantage of the conditioning-based solution over the
auxiliary-loss solutions may be attributed to the nature of the
conditioning operation in overhauling the internal structure of the
networks, where auxiliary-loss based solutions work mostly to provide
an external guidance. In other words, the former may have enhanced the
baseline network more fundamentally. Nonetheless, these two proposed
strategies both demonstrate the ability to improve the baseline model,
achieving our design goals.
%

%\begin{table*}[htb]% [!ht]
%  \caption{SI-SNR and STOI for different methods on LibriMix dataset.}
%\vspace{0.05in}
%\centering
%\scalebox{1.2}{
%\begin{tabular}{c|c|c|c}
%  \hlineB{2.5}

%{\textbf{Model}} &
%{\textbf{Auxiliary loss}} &
%{\textbf{SI-SNR (dB)}} &
%{\textbf{STOI}}\\

%\hlineB{2.5}
%         {Conv-TasNet} & {-} & {14.118} & {0.928}  \\ \hlineB{2.5}
%       {Conv-TasNet} & {weighted-sum} & {14.319} & {0.930}  \\ \hlineB{2.5}
% {Conv-TasNet} & {triplet-like} & {14.298} & {0.929}  \\ \hlineB{2.5}
% {Conditional} & {-} & {14.863}  & {0.936} \\ \hlineB{2.5}
% 

%\end{tabular}
%}
%\label{T:librimix}
%\end{table*}

\begin{table}[htb]% [!ht]
  \caption{SI-SNR and STOI for different methods on LibriMix dataset.}
\vspace{0.05in}
\centering
\scalebox{1.2}{
\begin{tabular}{c|c|c}
  %\hlineB{2.5} 
  %\hlineB{2.5}

  \hline
  \hline
  
{\textbf{Model}} &
{\textbf{SI-SNR (dB)}} &
{\textbf{STOI}}\\

  \hline
% \hlineB{2.5}
      {Conv-TasNet}  & {14.118} & {0.928}  \\ \hline  \hline 
      % \hlineB{2.5}   \hlineB{2.5}
        {$P_{\textrm{weighted-sum}}$} & {14.319} & {0.930}  \\ \hline % \hlineB{2.5}
 {$P_{\textrm{triplet-like}}$} & {14.298} & {0.929}  \\ \hline % \hlineB{2.5}
 {Conditioning} & {{\bf 14.863}}  & {{\bf 0.936}} \\ \hline % \hlineB{2.5}
 
\end{tabular}
}
\label{T:librimix}
\end{table}

The primary innovation in our proposed auxiliary-loss models lies in
the design and integration of $D_\textrm{neg}$, which aims to suppress
sounds of non-target speakers in the outputs. To investigate the
effect of this term, as well as the  relationship with the positive
distances $D_\textrm{pos}$, we conduct an ablation study upon the
proposed $P_\textrm{weighted-sum}$ model.
%
%As mentioned above, the positive distance
%$D_\textrm{pos}$ ensures separated voices sound like target speakers,
%and the negative distance $D_\textrm{neg}$ suppresses sounds of
%non-target speakers. While the perceptual loss in
%$P_\textrm{weighted-sum}$ is a weighted summation of $D_\textrm{pos}$
%and $D_\textrm{neg}$, these two models only use $D_\textrm{pos}$ and
%$D_\textrm{neg}$, respectively, in their auxiliary loss functions.
%
To this end, we implemented two additional models, which have the same
architecture as $P_\textrm{weighted-sum}$ but have only one part of
the proposed auxiliary loss. More specifically, the first model has a
loss function combining the basic loss (same as in Conv-TasNet) and
$D_\textrm{pos}$. The second model goes with the combination of basic
loss and $D_\textrm{neg}$. The baseline Conv-TasNet has only the basic
loss, and $P_\textrm{weighted-sum}$ can be regarded as a model with
basic loss + $D_\textrm{pos}$ + $D_\textrm{neg}$.

% We conduct an ablation study on $P_\textrm{weighted-sum}$ model to
% study the contributions of the positive and negative losses.  We
% compare our proposed weighted-sum perceptual loss based solution with
% two other models. These two models have the same architecture as our
% perceptual-loss based model, except that their auxiliary losses are
%designed differently. As mentioned above, the positive distance
% $D_\textrm{pos}$ ensures separated voices sound like target speakers,
% and the negative distance $D_\textrm{neg}$ suppresses sounds of
% non-target speakers. While the perceptual loss in
% $P_\textrm{weighted-sum}$ is a weighted summation of $D_\textrm{pos}$
% and $D_\textrm{neg}$, these two models only use $D_\textrm{pos}$ and
% $D_\textrm{neg}$, respectively, in their auxiliary loss functions.

The results are shown in Table \ref{T:ablation}.  While the model with
the $D_\textrm{pos}$ distance outperforms the baseline model,
performance of the former falls short of $P_\textrm{weighted-sum}$,
the model with combined three losses.  In contrast, performance of
``Basic loss + $D_\textrm{neg}$'' model is very close (but inferior)
to our combined model. These results indicate that the proposed
negative distance $D_\textrm{neg}$ plays a significant role in the
combined models, and is rather effective in reducing the residual
sounds of non-target speakers.

\begin{table}[t]% [!ht]
  % \caption{Ablation on LibriMix for the perceptual-loss based solution.}
    \caption{Ablation study on $P_\textrm{weighted-sum}$ for the
      effects of the positive and negative distances.}
\vspace{0.05in}
\centering
\scalebox{1.2}{
\begin{tabular}{c|c|c}
  % \hlineB{2.5} \hlineB{2.5}

  \hline \hline

{\textbf{Model}} &
{\textbf{SI-SNR (dB)}} &
{\textbf{STOI}}\\

\hline
       {Basic loss (Conv-TasNet)} & {14.118} & {0.928}  \\ \hline \hline
 {Basic loss + $D_\textrm{pos}$} &  {14.205} & {0.929}  \\ \hline
 {Basic loss + $D_\textrm{neg}$} &  {14.318}  & {0.929} \\ \hline
 {Basic loss + $D_\textrm{pos}$ +  $D_\textrm{neg}$} & {14.319} & {0.930} \\ \hline 
 %{RawNet2 features} &  {14.200}  & {0.928} \\ \hlineB{2.5}

%\hlineB{1.2}
%       {baseline} & {14.118} & {0.928}  \\ \hlineB{1.2}
% {only $L_1$ distance} &  {14.205} & {0.929}  \\ \hlineB{1.2}
% {only $L_2$ distance} &  {14.318}  & {0.929} \\ \hlineB{1.2}
% {perceptual (weighted-sum)} & {14.319} & {0.930} \\ \hlineB{1.2} 
% %{RawNet2 features} &  {14.200}  & {0.928} \\ \hlineB{2.5}

\end{tabular}
}
\label{T:ablation}
\end{table}

\section{Conclusion}

% In this paper, we propose a knowledge-assisted framework to enhance
% the perceptual properties of the denoised outputs of waveform SE
% networks.  Our approach relies on speech representations trained on
% large speech datasets to provide valuable insights and guidance
% regarding what clean speeches sound like. Experiments on both Noisy
% VCTK and TIMIT with SSNs datasets show that our models achieve
% significant perceptual performance gains to both the baseline and
% state-of-the-art models. The take-home message is that pretrained
% speech representation models, if properly integrated, do provide
% great help for SE. To explore integration of pretrained models with
% more speech enhancement networks, as well as their applications to
% other speech tasks, are our ongoing efforts.

In this paper, we propose two speaker-aware approaches to improve the
existing speaker separation solutions. The first strategy is to
integrate the information of speech samples to provide individualized
conditions for the separation module. Such individualization is
achieved through the combination of a basic model (Conv-TasNet) and a
pretrained SI network (RawNet2).

The second model falls in the auxiliary-loss based category. We design
{\it negative distances} to reduce the residual sounds from non-target
speakers and {\it positive distances} to strengthen target
outputs. Two different integration setups are design to combine the
proposed distances.
%Like most of speaker-conditioned models, this model
%comprises a speaker module and a separation module. In the speaker
%module, a pretrained SI model (RawNet2) takes separated signals out of
%a basic model (Conv-TasNet) as inputs and infers speaker embeddings
%for them. The inferred speaker embeddings is then taken as conditions
%to guide the process in the separation module. With this setting,
%speech information is integrated into our model because speaker
%embeddings produced by a SI model contains rich speech elements.
%
%Lastly, we design a novel auxiliary-loss based on SSL speech
%representations for the speaker separation task. The auxiliary loss
%functions of current solutions usually only have a term to ensure
%separated voices sound like target speakers. As a result, the residual
%sounds of non-target speakers can be clearly perceivable in the
%separated voices of these solutions. The motivation of this work,
%then, is to suppress these residual sounds. In implementation, the
%suppression is achieved by adding an term into the current auxiliary
%loss that maximizes the distance between speech representations of the
%predicted sources and the clean non-target sources. Two approaches are
%explored to combine this additional term with the original term in
%current auxiliary loss functions. The first approach applies a direct
%weighted sum of these two terms. In the second approach, we borrow the
%idea of the triplet loss and set a minimal margin between these two
%terms.
%
Experiments show the effectiveness of our proposed solutions. Exploring
more pretrained speech representation models, as well as studying their
guiding capabilities, is our ongoing effort.
%
% Our designed experiments show that this proposed model significantly
% outperforms the baseline Conv-TasNet model. In the next step, more
% pretrained DNN models that produce supervised speech feature maps or
% speech representations, such as those used in Chapter 2 to analyze the
% problem of guiding capabilities, will be explored to replace
% RawNet2.

\large
\bibliographystyle{IEEEtran}
\bibliography{speech}

\end{document}